\documentclass[aps,pra,showpacs,showkeys,twocolumn,superscriptaddress,floats,floatfix,preprintnumbers]{revtex4}
\usepackage{amssymb,amsmath}
\usepackage{bm,url,graphicx,subfigure,epsfig}
\graphicspath{{Pictures/}} 
\usepackage{color}
\usepackage{natbib}
\expandafter\let\csname equation*\endcsname\relax
\expandafter\let\csname endequation*\endcsname\relax
\usepackage{amsmath}
\usepackage{siunitx}

\usepackage{bm,graphicx,subfigure,epsfig}
\graphicspath{{Pictures/}} 
\usepackage{color}
\usepackage{pxfonts}
\usepackage[T1]{fontenc}

\begin{document}

\title{Landau-like states in neutral particles}
\author{Saikat Banerjee}
\affiliation{Institute of Materials Science, Los Alamos National Laboratory, Los Alamos, New Mexico, 87545, USA}
\affiliation{Nordita, Center for Quantum Materials, KTH Royal Institute of Technology and Stockholm University,
Roslagstullsbacken 23, 10691 Stockholm, Sweden}
\affiliation{Division of Theoretical Chemistry and Biology, Royal Institute of Technology, SE-10691 Stockholm,Sweden}
\author{Hans \r{A}gren}
\affiliation{Division of Theoretical Chemistry and Biology, Royal Institute of Technology, SE-10691 Stockholm,Sweden}
\author{A.V. Balatsky}
\affiliation{Institute of Materials Science, Los Alamos National Laboratory, Los Alamos, New Mexico, 87545, USA}
\affiliation{Nordita, Center for Quantum Materials,  KTH Royal Institute of Technology and Stockholm University,
Roslagstullsbacken 23, 10691 Stockholm, Sweden}
\date{\today,~  }

\begin{abstract}
We show the emergence of a new type of dispersion relation for neutral atoms with an interesting similarity with the spectrum
of 2-dimensional electrons in an applied perpendicular constant magnetic field. These neutral atoms can be confined in toroidal optical traps and give quasi Landau spectra. 
In strong contrast to the equi-distant infinitely degenerate Landau levels for charged particles, the spectral gap for such 2-dimensional neutral particles increases 
in particular electric field configurations. The idea in the paper is motivated by the development in cold atom experiments and builds on the seminal paper of Aharonov and Casher. 
\end{abstract}
\maketitle

\section{Introduction}

It is a well known that charged particles, like electrons, when subjected in two dimension to an applied perpendicular constant magnetic field generate Landau levels. 
The spectrum of these particles is infinitely degenerate. This happens because of the minimal coupling of the electromagnetic gauge field $ \bf A(\bf x) $ to 
the momenta of the particles in the 2-dimensional Hamiltonian. Landau levels are generated due to the interference of orbital motion of charged particles 
in the external magnetic field with a spectrum that is controlled by the magnitude of the physically observed field.  A similar question 
on the possibility of Landau levels generated by neutral particles can be asked. Following the same logic we need to find a field 
that would couple to the orbital motion of neutral particles. A natural possibility would be to consider the spin-orbit coupling of these particles. 

In this paper, we address the  question of formation of discrete Landau like  states for neutral atoms. In addition to similarities, the spin-orbit nature 
of the  coupling indicates clear differences for these states compared to the normal electronic Landau level problem as we explain below. 
According to the celebrated result from 1984  of Aharonov and Casher~\cite{Aharo}, we know that neutral particles with magnetic moment exhibits the 
Aharonov-Bohm effect under certain circumstances.  Due to relativistic effect a particle moving with velocity $\bf{v}$ in an electric field $\bf{E}$ will 
feel an effective magnetic field ${\bf{B}}=-({\bf{v \times E}})/c^{2}$~\cite{Eric}. Therefore, theoretically in the non-relativistic limit, one can write down 
a Rashba type interaction between a neutral particle with a magnetic moment and an electric field $\bf{E}$. The Hamiltonian for the system is given by~\cite{Bakke},

\begin{equation}\label{eq:1}
 \mathcal{H}=\frac{\bm{p}^2}{2m} +\alpha \bm{\sigma} \cdot (\bm{p} \times \bm{E})
\end{equation}

 In Eq.~(\ref{eq:1}) $\bf p$ is the momentum of the particle in two dimensions (assuming $xy$ plane), $m$ is its mass, $\bf E$ is the applied electric field 
and $\alpha$ is related to the magnetic moment as $\alpha \approx \frac{g\mu_{B}}{2mc^{2}}$, where $c$ is the speed of light, $g$ is the 
Lande-g factor and $\mu_{B}$ is the Bohr magneton.

 There are several earlier works on the interaction between neutral atoms and magnetic field which are of relevance to the present one - 
Paul and Philips {\it et. al.} discussed trapping of
neutral atoms~\cite{Paul},\cite{Migdall}; Schmiedmayer discussed the  trapping neutral atoms along a wire~\cite{Schm}; Ribeiro {\it et. al} 
analysed the Landau quantization of electric dipoles in the presence of crossed electric and magnetic fields~\cite{Furtado}; 
Spielman {\it et. al}~\cite{Spiel} discussed synthetic gauge fields in cold atomic spin-orbit (SO) physics. 
Motivated by the recent advances on cold atom physics, here we investigate the effect of synthetic electric fields on neutral atoms~\cite{Porto}.

 The outline of this paper is as follows. In Sec.~2, we explain the basic model and experimental setup for the interaction of neutral atoms 
and electric (synthetic) fields and its utilization to derive Landau like states for the atoms. In Sec.~3, we outline some realistic estimates 
of the typical Landau gaps and some other parameters of the model in context of cold atom physics. In the concluding Sec.~4, we discuss our 
work and also outline some possible future works based on this idea.
 
\section{Neutral atoms in synthetic electric field} 

 In this section, we describe the physics of the interaction between the neutral atoms and an effective electric field and how to utilize this interaction to find Landau like states for the atoms. 
We rewrite the Hamiltonian in Eq. (\ref{eq:1}) as,
 
\begin{subequations}
\begin{align}
\mathcal{H} &=\frac{\bm p ^2}{2m} +\alpha (\bm \sigma \times \bm p).\bm E 
\nonumber\\ &
   = \frac{\bm p ^2}{2m} + \alpha \bm p. \bm A_{eff} \quad \label{subeq1} \\
   \bm A_{eff} &= {\bm E}\times \bm \sigma \label{subeq2}
\end{align}
\end{subequations}

We see an analogue of the 'gauge' field coupling to the momentum of the neutral particles compared to the electromagnetic gauge coupling to the momentum of the charged particles~\cite{Jacob}. This "Gauge" field is related to the physical electric field~\cite{Lin}. We know that the Pauli matrices form an SU(2) representation. Therefore, the components of this gauge field (Eq.~(\ref{subeq1}) also forms an SU(2) representation~\cite{Mineev},\cite{Leurs},\cite{Estienne}. These are non-abelian gauge fields.
We assume an electric field $\bf{E}$ applied in the $\bf{\hat{y}}$ direction in the 2-Dimensional $xy$ plane of the form $\bm E=\gamma y^{2} \hat{\bm y}$ with a linear charge distribution. The specific choice 
of the electric field as $\bm E=\gamma y^{2} \hat{\bm y}$, which couple to the magnetic moments of the neutral particle, can be contrasted to the Landau level physics for charged particles. 
 
Recollecting the problem of Landau quantization in electrons, a constant magnetic field is obtained for a specific electromagnetic gauge as $A_x=By$ (considering a 2-dimensional electron gas). The magnetic 
moments of the neutral particles couple to the electric field similar to the minimal coupling of electromagnetic gauges for charged particles. The Hamiltonian in Eq. (\ref{eq:1}) 
becomes, (See Eq. (\ref{eq:7}) in Appendix. A)
  
\begin{equation}\label{eq:3}
  \mathcal{H}^{\pm}=\frac{p_{x}^2}{2m}\,+\frac{p_{y}^2}{2m}\, \pm \alpha \,\gamma\, y^{2}\, p_{x}
\end{equation}
 
We see that the Hamiltonian in Eq. (\ref{eq:3}) separates into two branches $\mathcal{H}^{\pm}$ considering the eigenvalues of $\sigma_{z}$ and does not depend on $x$. As we proceed with the 
simplification of the above Hamiltonian, we use $p_x$ as a good quantum number, which in a way will give Landau likes states. We can express the wave function as 
$\psi\, (x,y) = e^{i\,k_x\,x} \,\phi (y)$. Inserting $\psi\, (x,y)$ in Eq.~3, we get an effective one-dimensional Hamiltonian which acts on $\phi \,(y)$;

\begin{subequations} 
\begin{align}
{H^{\pm}_{eff}} &= -\frac{\hbar^2}{2m} \triangledown_{y}^2 \pm \alpha \, \gamma  \, \hbar\, k_x \, y^{2}{\label{subeq:3}} \\
& =-\frac{\hbar^2}{2m} \triangledown_{y}^2 \pm \frac{1}{2}m\, \omega_c^{2} y^{2} {\label{subeq:4}}
\end{align}         
\end{subequations}  
   
In Eq. (\ref{subeq:4}) the cyclotron frequency is $\omega_c^{2} =\frac{2\alpha \gamma \, \hbar\, k_x}{m}$. It depends on the mass of the neutral atoms as well as the on the $x$ component of the momentum, $k_x$. 
Only the $\sigma_z=+1$ branch of the Hamiltonian in Eq. (\ref{subeq:4}) gives the bound potential, $\frac{1}{2}m\, \omega_c^{2} y^{2}$, and the other branch $\sigma_z=-1$ is scattered and remains non-confined.
 
 \begin{figure}[!htb]
   \centering
   \includegraphics[width= .45\textwidth]{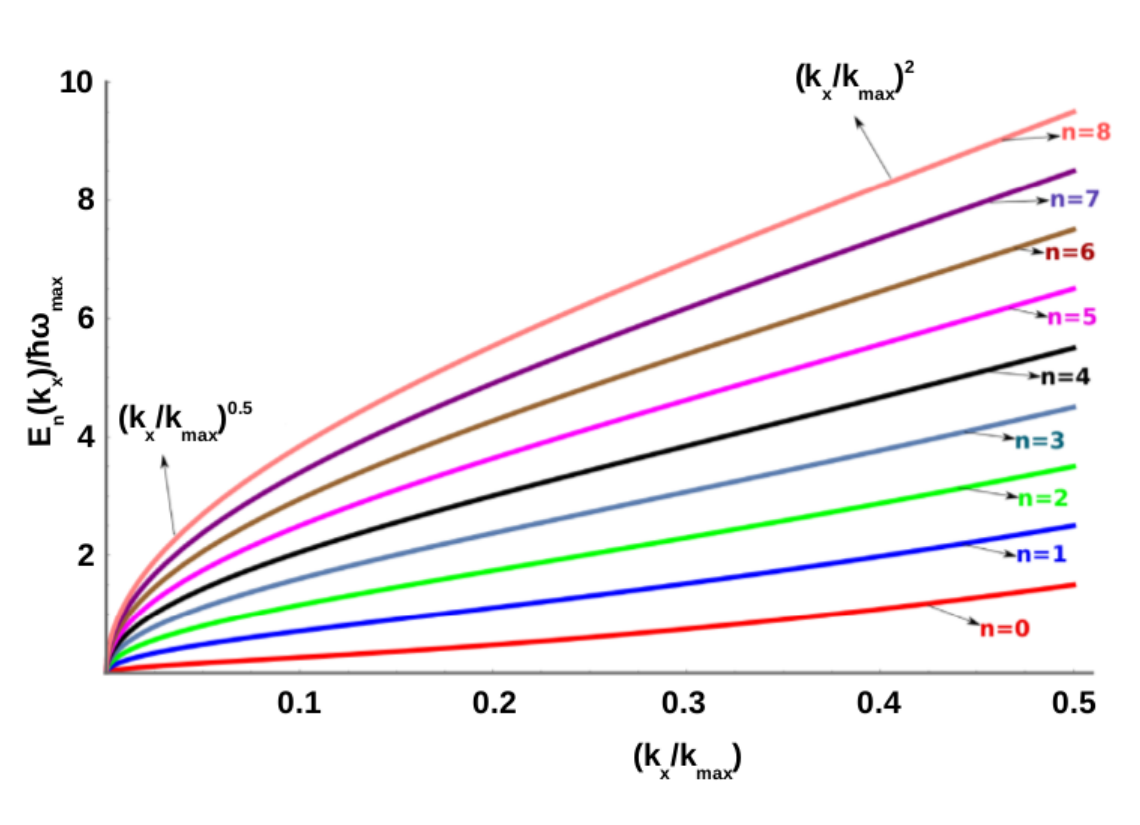}
   \caption{(Color Online) The plot of the spectrum [Eq. (\ref{subeq:6})] of the neutral atoms as a function of the momentum along $x$-direction. In contrast to the Landau levels of the 
charged particles, which are infinitely degenerate and equally spaced, the Landau like spectrum of the neutral particles, is non-degenerate and the non equi-spaced but similar to the standard Landau levels.}
   \label{Fig: Dispersion}
\end{figure}  

 We find the wave-function and the dispersion of the $\sigma_{z}=+1$ branch as
 
\begin{subequations}
\begin{align}
\psi_{n}(x,y) &= e^{ik_x x}\frac{1}{\sqrt{2^n n!}} \left(\frac{m\omega_c}{\pi \hbar}\right)^{1/4} e^{-\frac{m\omega_c}{2\hbar}y{^2}} H_{n}\left(\sqrt{\frac{m\omega_c}{\hbar}}y\right), {\label{subeq:5}}\\
  E_{n} &=\hbar \omega_c\left(n+\frac{1}{2}\right)+\frac{\hbar^2k_x{^2}}{2m}{\label{subeq:6}}
\end{align}
\end{subequations}
 
 The function $H_{n}(x)$ defined in Eq. (\ref{subeq:5}) is a Hermite polynomial and we find the dispersion relation $E_{n}$ in Eq. (\ref{subeq:6}) to be similar to that for the standard Landau levels. 
In contrast to the standard landau levels of the charged particle, which are infinitely degenerate, the dispersion relation in Eq. (\ref{subeq:6}) is not degenerate but depends on the momentum $\hbar k_{x}$, 
the spectrum is plotted in Fig. (\ref{Fig: Dispersion}). It is important to realize that only half of the particles will produce the landau-like spectra and the other half of the particles will scatter and 
behave like free particles. 

\section{Realization of the Landau physics of neutral atoms in optical traps} 
 
  In this section, we describe the physical background for realizing the Landau like spectra (Fig.~\ref{Fig: Dispersion}) for neutral particles. We consider neutral atoms in laser driven optical traps~\cite{Wineland}. 
Several laser beams are focused to interfere and create potential profiles such that the neutral atoms can reside near the minima of these potentials and create a quantum many-body system of ultra cold atoms. 
Our motivation to incorporate cold atoms in optical traps is that the analogy between the pseudo 'spins' and magnetic moments for the charge-less particles can be used~\cite{Goldman}. Consequently, we 
can ask if we can get similar type of spin-orbit (SO) coupling as in Eq.~(\ref{eq:1}) in this ultra cold many-body system~\cite{Lin}. We actually can have a Rashba type SO coupling by coupling the 
internal or $'$Dressed$'$ states of the cold neutral atoms with the laser field~\cite{Galitski},\cite{Larson},\cite{Neuman}. The coupling strength of these $'$Dressed$'$ states, 
in comparison with $\alpha$ in Eq. (\ref{eq:1}), depends on the laser profile. Hence, we use a pseudo spin-orbit interaction with trapped neutral atoms. 
The coupling constant $\alpha$ relates these pseudo 'spins' to the magnetic moments of the neutral particles~\cite{Ribeiro}. We know that for optical traps of cold atoms (neutral) one can talk about 
a synthetic electric field which couples to the 'pseudo' spins or dressed states of the neutral atoms~\cite{Porto}. The idea is to generate a synthetic quadratic electric field $\bf{E^{*}}$ of the 
form as discussed in Sec.~2. Coupling of this field to the dressed states of the atoms in the trap will generate the Landau like physics as we proposed in the previous section. Carefully examining the Hamiltonian 
in Eq.~(\ref{eq:3}) we see that both $\sigma_{z}=1$, $k_{x} > 0$ and $\sigma_{z}=-1$, $k_{x} < 0$ channel particles satisfy the harmonic oscillator Hamiltonian $\mathcal{H^{+}}$. However, there is a subtlety 
with realizing our model in realistic 2-dimensional optical traps. In a finite trap, the particles will be reflected from the boundary and change their momentum direction. Half of the
particles following $\mathcal{H^{+}}$ channel in Eq.~(\ref{eq:3}) will also be scattered when reflected from the boundary, as they will change the sign of the 
momentum $k_{x}$ (positive $k_{x}$ channel particles will become negative $k_{x}$ channel particles).  
  
\begin{figure}[!htb]
 \centering
   \includegraphics[width= .45\textwidth]{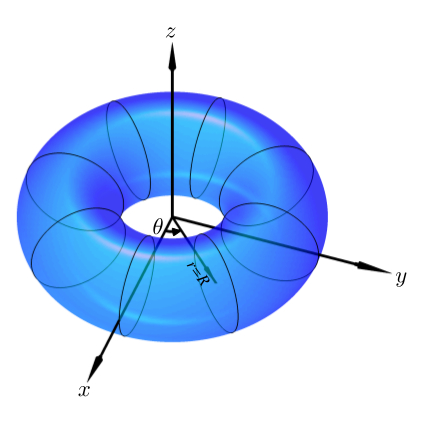}
   \caption{(Color Online) A schematic of toroidal potential trap used for realizing the Landau physics for neutral particles is shown in the above figure. The toroidal traps are incorporated 
to prevent the loss of particles from the trap due to reflection from boundary (See Eq.(\ref{eq:3})). For simplicity only the schematic of a typical toroidal potential trap is shown, exclusive of 
the laser sources used to create the trap and also the necessary "Spin-orbit" coupling for the Landau Physics (Eq.(\ref{subeq:6})).}
   \label{Fig: Toroid}
\end{figure}

  The most conducive way to prevent the loss of the particles from the trap is to incorporate toroidal optical traps~\cite{Lamb},\cite{Wright}. In such traps there will be no reflections because of the 
absence of any boundary and the particles following $\mathcal{H^{+}}$ branch will always remain in this branch and we can realize the Landau physics for these neutral atoms. A quasi two-dimensional 
toroidal optical trap is shown in Fig. (\ref{Fig: Toroid}).
  
We also consider the effect of temperature within the traps. The finite temperature of the traps will result in a momentum distribution for the particles. We may assume a Bose-Einstein distribution for 
these particles as $\frac{1}{exp \left(\beta ({\bf{p}}^{2}/2m) \right)-1}$. Therefore, we will see smeared out Landau spectra for these particles although the temperature in an cold atomic optical trap is 
close to absolute zero. We assume fermionic atoms (viz. $^{40}K$)~\cite{Ridinger},\cite{McKay} in the traps so that they do not form any Bose-Einstein condensate (BEC). We therefore exclude the 
effects of vortices which will be important if there are BEC is in the traps\cite{Spiel}. The spin-orbit coupling for the neutral particles in the Hamiltonian [Eq.~(\ref{eq:3}) and (\ref{eq:7})] is an 
example of equal admixtures of Rashba and Dresselhaus interactions in standard solid state physics~\cite{Dresselhaus}. Generalized Rashba-Dresselhaus SO coupling can be generated in 
the cold atomic traps~\cite{Juzeli}. Using accurate laser profiles and couplings we can generate a similar coupling as in Eq.~(\ref{eq:6}). 
 
\section{Parameter tuning for Landau physics of neutral atoms in optical traps}
 
 In this section, we discuss the tunability of the parameters of our model and whether it is possible for experimentally verify our proposal. A typical energy gap is of the order of $\hbar \omega_c$ which 
is $\hbar \sqrt{(2\alpha \gamma \, v_x)}$. As discussed earlier, assuming the neutral particles inside the trap will satisfy a Boltzmann distribution, there will be a mean speed for the 
particles corresponding to the temperature of the trap. We estimate the typical level spacing for the Landau like states by assuming some realistic values of the parameter for our theoretical model 
as $\gamma =10^{10} V/m^3$ and $ v_x=10^{-1} m/s $ and $\alpha =3.6 \times 10^{-16} m^2 s^{-1}/V $ of the order of $ 5.504 \times 10^{-19} eV $. The estimate for $\alpha$ is calculated from 
atomic masses of the relevant atoms and the magnetic moment of the hyperfine spin state and speed of light $c$. We see that the gap is extremely small. It is impossible to measure such small gaps 
with current technology. In contrast, in atomic traps we can produce the synthetic electric field which can couple to the dressed states of the trapped atoms as discussed in Sec.~3. If we can 
produce a similar synthetic electric field with a SO coupling as $\mathcal{H}_{so}=\beta y^{2}k_{x}\sigma_{z}$ we can use the freedom of tuning the coupling constant $\beta$ to get a
substantial measurable gap in the spectrum (Eq.~(\ref{subeq:6})). The typical energy gap in this case is $\hbar \omega_{c}'=\hbar \sqrt{(2\beta v_{x})/\hbar} $, where the cyclotron frequency 
$\omega_{c}'^{2}=(2\beta k_x)/m$. If we choose $\beta \approx 10^{-20} \si{kg.m/s^{2}}$ \cite{Goldman}, the level spacing between $n=0$ and $n=1$ level for the same mean speed 
is approximately $2.8\times 10^{-9}$ eV. The corresponding temperature scale is of the order of few microKelvin $ 2.8\times 10^{-9} eV \sim 3.25\times 10^{-5} K$ which is attainable in 
optical traps. Therefore, we may observe neutral atoms Landau levels in optical traps~\cite{Duarte}.

\section{Discussion and Conclusion}

In  this paper we demonstrated that upon the application of quadratic electric field and induced spin-orbit coupling in optically trapped cold
atoms, one can induce non-trivial spin dependent  levels in the spectrum of the atoms. In contrast to normal Landau levels where the energy spectrum is 
equispaced, we find here  a different energy spectrum with continuously increasing level spacing as a function of the phase momenta. We note that a new type of gauge field arises and couples 
to the momentum of the neutral particles with magnetic moments in the presence of specific charge distribution. This is to be contrasted with the  Landau level physics of electrons in applied magnetic fields. 
The scattering of half of the particles in the optical trap due to the presence of $\sigma_{z}$ in the Hamiltonian in Eq. (\ref{eq:3}) can be prevented with toroidal optical traps as 
presented in Fig. (\ref{Fig: Toroid}).

We have given an order of magnitude estimate of the gaps in the spectrum with some reasonable values of the parameters. We also keep in mind the technical difficulty of creating a spatial spin-orbit coupling 
with gradient Zeeman splitting. If we choose the synthetic SO coupling constant $\beta$ as small as  $10^{-26}\si{kg.m/s^{2}}$, the corresponding temperature scale for the level spacing at the same mean speed is approximately $10^{-9}$ K.  We see that the range of the level spacing is from $\mu$K to nano-Kelvin for a range of choice of the value of the coupling constant $\beta$. This level spacing can be measured experimentally and it makes our proposal more prone to 
practical verification. Another important point to note is that the cyclotron frequency depends on the x-component of the phase momentum $k_{x}$. For big wave numbers one can therefore tune the energy gap 
to any value that is practical. At same time there are experimental limitations for observing a very large $k_x$. The cold atomic experiments are performed at a very low temperature and so these atoms can not have a large momentum.

While we considered the trapped atoms being Fermionic like $^{40}K$, our approach is still valid for non-BEC bosonic system, for example; if  one takes $^{87}Rb$ the transition temperature for 
the gas cloud to form Bose-Einstein condensates is approximately $0.2$ Micro-kelvin. In our estimate, the temperature range is of the order of $10$ Micro-kelvin. In that case, the atoms do not 
condense and a simple quasi-Landau spectrum can be found.
Ultimately the proposed mechanism to control neutral particles might be useful for optical applications and indicates the possibility to  produce a Hall effect in neutral particles.

\section{Acknowledgment}

SB is grateful to Jonas Larson, Stockholm University and Erik Sj\"{o}qvist, Uppsala University for many important discussions regarding the idea of the paper and cold atomic experiments. This work was supported by ERC DM 321031 and US DOE BES E3B7 and by the Swedish Science Research Council (Grant 2012-3447). SB and H{\AA} acknowledge The Knut and Alice Wallenberg foundation for financial support (Grant No. KAW-2013.0020).

\appendix
\section{Spin-Orbit Hamiltonian}
   
 In this appendix we show the mathematical derivation of two branched Hamiltonian discussed in Sec.~2 in Eq.~(\ref{eq:3}). Taking the particular form of the Electric field ${\bf{E}}=\gamma y^{2}\hat{\bm{y}}$ as discussed in Sec.~2, the Hamiltonian in Eq.~(\ref{eq:1}) simplifies as follows,

\begin{align}{\label{eq:6}}
\nonumber
H &=\frac{\bm p ^2}{2m} +\alpha (\bm \sigma \times \bm p).\bm E \\
\nonumber
&=\frac{\bm p ^2}{2m} +\alpha \gamma y^{2} \hat{\bm y}. (\bm \sigma \times \bm p)\\
\nonumber
&=\frac{\bm p ^2}{2m} +\alpha \gamma y^{2} {\bm p}.(\hat{\bm y} \times \bm \sigma )\\
\nonumber
&=\frac{\bm p ^2}{2m} +\alpha \gamma y^{2} (p_{x} \hat{\bm{x}}+p_{y}\hat{\bm{y}}).(-\sigma_{x} \, \hat{\bm z}+\sigma_{z} \, \hat{\bm x} )\\
\nonumber
&=\frac{\bm p ^2}{2m} +\alpha \gamma y^{2} p_{x} \sigma_{z} \\
&=\frac{p_{x}^2}{2m}+\frac{p_{y}^2}{2m} + \alpha \gamma y^{2} p_{x} \sigma_{z} 
\end{align}

 The Pauli matrix $\sigma_{z}$ has two eigenvalues. Hence, the Hamiltonian separates into two branches $\mathcal{H}^{\pm}$ for $\sigma_{z}=\pm 1$ as,

\begin{equation}\label{eq:7}
  \mathcal{H}^{\pm}=\frac{p_{x}^2}{2m}\,+\frac{p_{y}^2}{2m}\, \pm \alpha \,\gamma\, y^{2}\, p_{x}
\end{equation}

\section*{References}

\bibliographystyle{iopart-num}
\bibliography{multimode}

\providecommand{\newblock}{}
\begin{thebibliography}{10}
\expandafter\ifx\csname url\endcsname\relax
  \def\url#1{{\tt #1}}\fi
\expandafter\ifx\csname urlprefix\endcsname\relax\def\urlprefix{URL }\fi
\providecommand{\eprint}[2][]{\url{#2}}

\bibitem{Aharo}
Aharonov Y and Casher A 1984 {\em Phys. Rev. Lett.\/} {\bf 53}(4) 319--321

\bibitem{Eric}
Ericsson M and Sj\"oqvist E 2001 {\em Phys. Rev. A\/} {\bf 65}(1) 013607

\bibitem{Bakke}
Bakke K and Furtado C 2009 {\em Phys. Rev. A\/} {\bf 80}(3) 032106

\bibitem{Paul}
Paul W 1990 {\em Rev. Mod. Phys.\/} {\bf 62}(3) 531--540

\bibitem{Migdall}
Migdall A~L, Prodan J~V, Phillips W~D, Bergeman T~H and Metcalf H~J 1985 {\em
  Phys. Rev. Lett.\/} {\bf 54}(24) 2596--2599

\bibitem{Schm}
Schmiedmayer J 1995 {\em Phys. Rev. A\/} {\bf 52}(1) R13--R16

\bibitem{Furtado}
Furtado C, Nascimento J and Ribeiro L 2006 {\em Physics Letters A\/} {\bf 358}
  336 -- 338 ISSN 0375-9601

\bibitem{Spiel}
Lin Y~J, Compton R~L, Jimenez-Garcia K, Porto J~V and Spielman I~B 2009 {\em
  Nature\/} {\bf 462}(7273) 628--632

\bibitem{Porto}
Lin Y~J, Compton R~L, Jimenez-Garcia K, Phillips W~D, Porto J~V and Spielman
  I~B 2011 {\em Nat Phys\/} {\bf 7}(7) 531--534

\bibitem{Jacob}
Jacob A, Ohberg P, Juzeliunas G and Santos L 2008 {\em New Journal of
  Physics\/} {\bf 10}(4) 045022

\bibitem{Lin}
Lin Y~J, Jimenez-Garcia K and Spielman I~B 2011 {\em Nature\/} {\bf 471} 83--86
  ISSN 0028-0836

\bibitem{Mineev}
Mineev V and Volovik G 1992 {\em Journal of Low Temperature Physics\/} {\bf 89}
  823--830 ISSN 0022-2291

\bibitem{Leurs}
Leurs B, Nazario Z, Santiago D and Zaanen J 2008 {\em Annals of Physics\/} {\bf
  323} 907 -- 945 ISSN 0003-4916

\bibitem{Estienne}
Estienne B, Haaker S~M and Schoutens K 2011 {\em New Journal of Physics\/} {\bf
  13}(4) 045012

\bibitem{Wineland}
Wineland D and Itano W 1979 {\em Phys. Rev. A\/} {\bf 20}(4) 1521--1540

\bibitem{Goldman}
Goldman N, Juzeliūnas G, Öhberg P and Spielman I 2014 {\em Reports on
  Progress in Physics\/} {\bf 77}(12) 1--60

\bibitem{Galitski}
Galitski V and Spielman I~B 2013 {\em Nature\/} {\bf 494} 49--54 ISSN 0028-0836

\bibitem{Larson}
Larson J, Martikainen J~P, Collin A and Sj\"oqvist E 2010 {\em Phys. Rev. A\/}
  {\bf 82}(4) 043620

\bibitem{Neuman}
Neuman K~C and Block S~M 2004 {\em The Review of scientific instruments\/} {\bf
  75} 2787--809 ISSN 0034-6748

\bibitem{Ribeiro}
Ribeiro L, Furtado C and Nascimento J 2006 {\em Physics Letters A\/} {\bf 348}
  135 -- 140 ISSN 0375-9601

\bibitem{Lamb}
Lembessis V~E, Ellinas D and Babiker M 2011 {\em Phys. Rev. A\/} {\bf 84}(4)
  043422

\bibitem{Wright}
Wright E~M, Arlt J and Dholakia K 2000 {\em Phys. Rev. A\/} {\bf 63}(1) 013608

\bibitem{Ridinger}
Ridinger A, Chaudhuri S, Salez T, Eismann U, Fernandes D, Magalhães K,
  Wilkowski D, Salomon C and Chevy F 2011 {\em The European Physical Journal
  D\/} {\bf 65} 223--242 ISSN 1434-6060

\bibitem{McKay}
McKay D, Jervis D, Fine D, Simpson-Porco J, Edge G and Thywissen J 2011 {\em
  Phys. Rev. A\/} {\bf 84}(6) 063420

\bibitem{Dresselhaus}
Dresselhaus G 1955 {\em Phys. Rev.\/} {\bf 100}(2) 580--586

\bibitem{Juzeli}
Juzeli\ifmmode~\bar{u}\else \={u}\fi{}nas G, Ruseckas J and Dalibard J 2010
  {\em Phys. Rev. A\/} {\bf 81}(5) 053403

\bibitem{Duarte}
Duarte P, Hart R, Hitchcock J, Corcovilos T, Yang T~L, Reed A and Hulet R 2011
  {\em Phys. Rev. A\/} {\bf 84}(6) 061406

\end{thebibliography}

\end{document}